\documentclass[conference,compsoc]{IEEEtran}

\ifCLASSOPTIONcompsoc
  \usepackage[nocompress]{cite}
\else
  \usepackage{cite}
\fi
\ifCLASSINFOpdf
  \usepackage[pdftex]{graphicx}
  \graphicspath{{./images/}}
  \DeclareGraphicsExtensions{.jpg}
\else
 \usepackage[dvips]{graphicx}
\fi
\usepackage{amsmath}
\usepackage{amssymb}

\usepackage{url}

\usepackage{lipsum}
\usepackage{dirtytalk}
\usepackage{subcaption}
\usepackage{amsfonts}

\usepackage[inline]{enumitem}

\usepackage{tikz}
\usetikzlibrary{calc,positioning,arrows}

\newcommand{\imageref}[4][0.5]{
  \begin{figure}
    \includegraphics[width=#1\textwidth]{#2}
    \caption{#3}
    \label{fig:#4}
  \end{figure}
}

\usepackage{xcolor}
\usepackage{listings}

\newcommand\JSONnumbervaluestyle{\color{red}}
\newcommand\JSONstringvaluestyle{\color{red}}

\newif\ifcolonfoundonthisline

\makeatletter

\lstdefinestyle{json}
{
  showstringspaces    = false,
  keywords            = {false,true},
  alsoletter          = 0123456789.,
  morestring          = [s]{"}{"},
  stringstyle         = \ifcolonfoundonthisline\JSONstringvaluestyle\fi,
  MoreSelectCharTable =%
    \lst@DefSaveDef{`:}\colon@json{\processColon@json},
  basicstyle          = \ttfamily,
  keywordstyle        = \ttfamily\bfseries,
}

\newcommand\processColon@json{%
  \colon@json%
  \ifnum\lst@mode=\lst@Pmode%
    \global\colonfoundonthislinetrue%
  \fi
}

\lst@AddToHook{Output}{%
  \ifcolonfoundonthisline%
    \ifnum\lst@mode=\lst@Pmode%
      \def\lst@thestyle{\JSONnumbervaluestyle}%
    \fi
  \fi
  \lsthk@DetectKeywords%
}

\lst@AddToHook{EOL}%
  {\global\colonfoundonthislinefalse}

\begin{document}

\IEEEoverridecommandlockouts
\IEEEpubid{\makebox[\columnwidth]{978-1-5090-3216-7/16/\$31.00~\copyright{}2016 European Union\hfill}\hspace{\columnsep}\makebox[\columnwidth]{ }}

%
\title{MANETs Monitoring with a Distributed Hybrid Architecture}

\author{

  \IEEEauthorblockN{Jose Alvarez and Stephane Maag}
  \IEEEauthorblockA{
    SAMOVAR, Telecom SudParis, Universit\'e Paris-Saclay\\
    9 Rue Charles Fourier, 91000, Evry, FR\\
    \{jose\_alfredo.alvarez\_aldana,\\
    stephane.maag\}@telecom-sudparis.eu
  }

\and

  \IEEEauthorblockN{Fatiha Za\"{i}di}
  \IEEEauthorblockA{
    LRI-CNRS, Universit\'{e} Paris Sud, Universit\'e Paris-Saclay\\
    15 Rue Georges Clemenceau, 91400, Orsay, FR\\
    Fatiha.Zaidi@lri.fr
  }
}

\maketitle

\begin{abstract}

Monitoring techniques have been deeply studied in wired networks using gossip and hierarchical approaches. 
However, when applied to a MANET, several problematics arise. 
We present a hybrid distributed monitoring architecture for MANETs. 
We get inspired of gossip-based and hierarchical-based algorithms for query dissemination and data aggregation. 
We define gossip-based mechanisms that help our virtual hierarchical topology to complete the data aggregation, and then ensure the stability and robustness of our approach in dynamic environments. 
We propose a fully distributed monitoring protocol that ease the nodes communications. 
We evaluate our approach 
by using NS3 and Docker. 

\end{abstract}


%
\IEEEpeerreviewmaketitle

\section{Introduction} \label{introduction}

Network monitoring have been deeply studied in P2P, DTN and others using gossip-based or hierarchical-based approaches. 
However, when it is applied to a wireless mobile ad hoc network (MANET), new problematics arise mainly due to the absence of a centralized administration, the inherent MANETs properties and the node mobility.
Some approaches propose a coordinator, nevertheless, due to energy efficiency, 
infrastructure or other parameters, these solutions are not always applicable.

While studying the monitoring of a network, the most common and intuitive approach is to define a central node as a coordinator for storage and processing of the observations.
This is notably proposed by \cite{cormode2013continuous}, where the author surveys the different communication mechanisms. 
These centralized architectures might be efficient for certain type of topologies, but become critical when considering dynamic topologies.
This is why there has been a lot of efforts on decentralized monitoring.
Gossip-based approaches show an extraordinary robustness and stability in dynamic scenarios and changing topologies.
Nonetheless, depending on the scalability, the cost and performance can be impacted.
On the other side, hierarchical approaches show an efficient performance, cost and scalability, although the  robustness and stability may decrease in dynamic scenarios.
This shows that the two major categories perform very good under different characteristics, requirements and constraints of a network \cite{stingl2012benchmarking}.

The main contribution of this paper is the proposal of a hybrid algorithm for decentralized monitoring of MANETs.
We define an architecture combining gossip-based and hierarchical-based algorithms for query dissemination and data aggregation.
We perform the gossip-based approach to disseminate the query and in the process to build a virtual hierarchical topology (VHT) for a time window.
Once the query is disseminated through all the network, with the support of the VHT, a hierarchical-based aggregation takes place.
The second contribution of this paper is the definition of a monitoring protocol that aims at helping a decentralized monitoring process.
Our expectation is not just to provide a structure but also a mathematical background for further model checking and testing. 
Our protocol has been successfully assessed using NS3 and Docker. 


The remaining of our paper is as it follows. 
In Section \ref{methods}, we present our hybrid algorithm.
In Section \ref{results}, we present our implementation, with a semi-formal support for our protocol. 
Next, in Section \ref{relatedworks}, we present some interesting related works from which we got inspired. 
Finally, we conclude and give some perspectives in Section \ref{conclusions}.



\section{Hybrid Monitoring Approach} \label{methods}

Network monitoring can be described as \say{A number of observers making observations, and wish to work together to compute a function of the combination of all their observations} \cite{cormode2013continuous}. 
The goal is that all the nodes in the network compute a value $t\mapsto f(t)$ [$\mathbb{R}^{+*}\rightarrow X$, $X$ being the domain targeted by $f$] in a given instant of time $t$ in a collaborative way.
For our purposes, $f$ is a linear and non-complex function (e.g., the average CPU).

Our hybrid algorithm architecture consists in two network states, the \say{query state} and the \say{aggregate state}.
The idea is to combine a gossip approach and a hierarchical approach to achieve the monitoring of a property of the network.
The communication between the nodes to achieve the monitoring of the network will be achieved through a package previously defined.
The idea is that a start node will start the monitoring process by propagating a monitoring query in a gossip approach.
The approach chosen will be described as epidemic. 
Each hop, the nodes will exchange information creating a virtual hierarchical topology (VHT) which will be valid only during the monitoring process.
Then based on this topology, the nodes will start aggregating the information by sending their results to the parent node.
Once the aggregation is done and has reached the start node, there will be a global view of the measured property and the VHT will no longer be usable.
If the process starts again, a new VHT will be derived.
The purpose is to establish a VHT during one time window, duration of the monitoring process, to ease the analysis and the global view of the property.

\subsection{Hybrid Architecture} \label{hybridArchitecture}

\subsubsection{Query State} \label{queryState}

The query state refers to the process of propagating in an epidemic way the monitoring packet.
This state goal is to disseminate the query and the VHT layout to allow the nodes in the network to do an accurate and efficient aggregation in the next state.
This query will be forwarded in an epidemic approach to the nodes in the relay set.
The packet is explained in depth in Section \ref{results}, containing the query itself but also the information to generate the VHT.
This will communicate all the network information to create the VHT, which is the foundation of the following state of the network.
This process will go on until a node on the edge of the network is reached.

Along this state, there are some specific challenges to discuss.
\begin{enumerate*}[label=(\roman*)]
  \item The first challenge is if a node receives more than one monitoring packet once it is already in a monitoring state.
  For this, the node will take the first monitoring packet and will discard all the subsequent monitoring packets.
  \item The second challenge is, what if the propagation of the query is interrupted by a node that remains in a cyclic state. 
  For this, we introduce a timeout for the packet to avoid these problems. 
  The idea is to provide a mechanism to avoid loops in the communications.
  For this problem, the timeout will be triggered and once reached, this node will start the aggregation process by sending its result to the parent node.
  \item The third challenge is the broadcast of the packet itself.
  Due to the nature of the simple epidemic dissemination approach, a packet will be forwarded to the next hop of nodes but also to the parent node. 
  We decided that this will work as an acknowledgment of the child node to the parent node.
  This way, the parent node will receive $n$ 
  acknowledgments and he will know how many packets he should wait for before changing to the aggregate state.
\end{enumerate*}

\subsubsection{Aggregate State} \label{aggregateState}

Once the data is disseminated up to the edge of the network, the edge nodes will change from query state to aggregate state and will start sending recursively their information to their parent up to the start node.
This process will be an aggregation of all the data of a node and his children in order to collect the monitored values.
The aggregation will be computed in a hierarchical manner with a combination when required of a gossip approach.
A node will compute based on his own observations the result of the function $f(x)$ that received from the query state.
This information will be aggregated with the same child nodes information.
In the edge nodes cases, where this state starts, it will be done only with information from themselves.

Along this state, there are some specific challenges to discuss.
\begin{enumerate*}[label=(\roman*)]
  \item The first challenge is when a parent node and a corresponding child node goes out of range from when they first met.
  When the child node sends an aggregate type of message and receives no acknowledgment it will trigger a forward packet to the corresponding node.
  For this, we will rely on the routing protocol of the network.
  This makes our approach dependent on the routing layer of the network and we will consider our own opportunistic routing mechanism in forthcoming works.
  \item The second challenge is when a parent node is off line.
  For this, we propose that in the query state, a set of nodes are communicated to every child node for them to have an alternative path.
  Since the child node will have the relay set of the parent node, he will fall back into one of these nodes to send the information.
  Since it is a hierarchical approach, the parent will send the information about his parents in the VHT.
  \item The third challenge is when a node receives a grandchild node aggregate information.
  For this, the node will assume that the child node is off line and that he will be aggregating that information.
  Given that the node does not know the information of how many grandchild will send information, he will also rely on the timeout before he sends his own aggregate information.
  For every grandchild packet he receives, he will restart the timeout to give time for additional packages.
  If the timeout is reached, he will continue with his aggregation process.
\end{enumerate*}


\section{Experiments} \label {results}


\subsection{Protocol Definition} \label{protocolAutomata}

\begin{figure}
\centering

\scalebox{0.50}{%
\begin{tikzpicture}[->,>=stealth',shorten >=1pt,auto,node distance=5cm,
  thick,main node/.style={circle,fill=blue!20,draw,
  font=\sffamily\small\bfseries,minimum size=8mm}]

  \node[main node] (I) {Initial};
  \node[main node] (Q1) [right of=I,node distance=3.85cm] {Q1};
  \node[main node] (Q2) [below right of=Q1] {Q2};
  \node[main node] (A1) [below of=Q2] {A1};
  \node[main node] (A2) [left of=A1,node distance=4.15cm] {A2};
  \node[main node] (A3) [left of=A2] {A3};

  \path[every node/.style={font=\sffamily\small,
      fill=white,inner sep=1pt}]

  (I) edge [] node[align=center] {$startMonitoring()$/\\$SND Query$\\\\$RCV Query$/\\$SND Query$} (Q1)

  (Q1) edge [] node[align=center] {$RCV QueryACK$/\\$acc(ACK\_IP)$} (Q2)
  (Q1) edge [bend right=25] node[align=center] [near end] {$timeout()$/\\$SND Aggregate$} (A1)

  (Q2) edge [loop left] node[align=center] {$RCV QueryACK$/\\$acc(ACK\_IP)$} (Q2)
  (Q2) edge [] node[align=center] [near end] {$RCV Aggregate$/\\$SND Aggregate$} (A1)

  (A1) edge [] node[align=center] {$timeout()$/\\$SND AggregateRoute$} (A2)

  (A2) edge [] node[align=center] {$timeout()$/\\$SND AggregateForward$} (A3)

  (A3) edge [loop left] node[align=center] {$timeout()$/\\$Aggregate Forward$} (A3)

  (A3) edge [bend left=30] node[align=center] {$RCV AggregateACK$/\\$done()$\\\\$emptyForwards()$/\\$error()$} (I)
  (A2) edge [bend left=25] node[align=center] {$RCV AggregateACK$/\\$done()$} (I)
  (A1) edge [bend left=20] node[align=center] [near end] {$RCV AggregateACK$/\\$done()$} (I);

\end{tikzpicture}
}

\caption{State machine definition of our protocol} \label{fig:automata}
\end{figure}

The protocol definition, depicted in Figure \ref{fig:automata}, shows the expected behavior of the protocol to support as base ground for the hybrid monitoring architecture.
The set of states is $Q = (Initial, Q1, Q2, A1, A3, A3)$.
Where $Initial$ is the initial state.
The states $Q1$ and $Q2$ refer to the query states of the network.
And the states $A1$, $A2$ and $A3$ refer to the aggregate states of the network.
The internal operations of the automaton are startMonitoring(), acc(IP), timeout(), done() and error().
The startMonitoring() refers to the process of starting the monitoring.
The acc(IP) refers to the process of the node of accumulating the IP of the acknowledgment messages source.
This is used to identify while the query is propagating if there are child nodes available for a given node.
If a node does not receive any acknowledgment, he will continue the monitoring process by using a timeout.
The timeout() refers to the process of counting time since the last package received.
The done() refers to 
the restart of the state of the node.
Meanwhile, error() refers to the process of not being able to send a message, which if it happens, it means that the node itself is out of the network range or a major outage is happening with the network.
The input and output operations of the automaton are determined by sending (SND) and receiving (RCV) messages.
The possible messages to be sent or received are the query, query ack, aggregate, aggregate ack, agregate route and aggregate forward.
The query message refers to the query itself and the base ground of the query state.
For simplicity purposes, in the automaton, there is a distinction between the query and the query ack message.
But in reality, they are meant to be the same package but received by a different node. 
This is discussed in Section \ref{queryState}.
The aggregate messages refer to the aggregation process and the same principle applies as the query messages.
The aggregate ack message is an aggregate message but received by a different node.
Then we also have two extra messages which are the aggregate route and aggregate forward.
The aggregate route message refers to the process of routing a message through the network to the corresponding parent node in the VHT. 
As explained in Section \ref{aggregateState}, the idea is to make the hierarchical aggregation more robust through the addition of a gossip routing approach to route the package to the corresponding root node on the fly.
And finally, the aggregate forward message, which in the case that the parent node is not found, probably because the parent node went offline due to an outage or something similar.
In this case, the message will be forwarded to one of the nodes defined in the relay set, which will be populated by the grandparents and siblings.

\subsection{Packet Definition} \label{packetDefinition}

In order for the communication to be successful, we need to define the monitoring packet.
The packet will work equally in both states of the network, query and aggregate states, but different information will be sent depending on the state containing a set of common properties. 
It needs to contain some basic information in order to be useful for the following nodes and hops.
The definition of such packet will be done using json. 
For each state of the nodes, there will be a set of properties transmitted.
There will be a set of global properties that will always be transmitted. 
These global properties are:
\begin{enumerate*}[leftmargin=0.4cm]
  \item Type: the type of message being sent, the set of values is listed in \ref{protocolAutomata}.
  \item Parent: the IP address of the parent node. 
  \item Source: the IP address of the node sending the message.
  \item Timeout: the timeout value in milliseconds.
\end{enumerate*}
For the query state the properties transmitted are: 
\begin{enumerate*}[leftmargin=0.4cm]
  \item Function: the function $f$ to compute. 
  \item Relay Set: the list of IPs for alternative paths, with at most three items. 
\end{enumerate*}
For the aggregate state the properties transmitted are:
\begin{enumerate*}[leftmargin=0.4cm]
  \item Result: the result of the aggregation of the function $f$. 
  \item Destination: the destination IP that should be the parent IP for most of the cases, unless the parent node is off line, then it will be a relay set IP.
  \item Observations: the number of aggregated observations. 
\end{enumerate*}
The json definition of the complete package is the following:

\begin{center}
{\tiny
\begin{lstlisting}[style=json]
{"type": "<type>", "parent": "<parent IP>", "source": "<source IP>", 
 "timeout": <ms>,
 "query":{ "function": "<f(t)>", "relaySet": ["<IP list relay set>"] },
 "aggregate":{ "outcome": "<monitoring result>", "destination": <destination IP>,
  "observations": <number of observations> } }
\end{lstlisting}
\vspace{-1em}
}
\end{center}

\subsection{Results} \label{implementation}

{\small
\begin{table}
  \caption{Scenario 1 \& 2 parameters}\label{table:scenario1param}
  \vspace{-1em}
  \begin{center}
      \begin{tabular}{| l | l | l |}
      \hline
       & \textbf{Scenario 1} & \textbf{Scenario 2} \\ \hline
      \textbf{Number of nodes} & \begin{tabular}[x]{@{}c@{}}10, 20, 25, 40, 50, \\ 60, 75, 80 and 100\end{tabular} & 25  \\ \hline
      \textbf{Network Space} & \begin{tabular}[x]{@{}c@{}}500x500, 800x800 \\ and 1,000x1,000\end{tabular} & 500x500  \\ \hline
      \textbf{Network Positioning} & Grid (100m apart) & Random  \\ \hline
      \textbf{Running time} & 80s (init time 60s) & 80s (init time 60s) \\ \hline
      \textbf{Emulation times} & 200 & 40  \\ \hline
      \textbf{Mobility} & - & RWP  \\ \hline
      \textbf{Mobility Speed} & - & 2m/s and 5m/s  \\ \hline
      \end{tabular}
  \end{center}
  \vspace{-2em}
\end{table}
}


We evaluate our proposal using an emulator built in-house based on DOCKEMU \cite{to2015dockemu}. 
This emulator is a combination between Docker and NS3, which allows to conduct highly scalable, replicable and robust experiments.
The testbed consisted in an implementation of the protocol in the language Go, that was deployed on our emulator.
The idea was to determine the convergence time, by which we mean the time it took from the moment that the monitoring started by the root node, to the moment that the root node was able to return a verdict.
We defined two scenarios, one scenario with no mobility and another with low mobility.
For this study, we are testing the implementation without the mobility support. This means we do not consider states $A2$ and $A3$ of Figure \ref{fig:automata}.
Scenario 1 and scenario 2 consider the parameters of Table \ref{table:scenario1param}.
For both scenarios the Mac Protocol is 802.11a with a data rate of 54Mbps.
Each node had a range of $\approx$125m. 
Scenario 1 was designed to test the convergence time, the scenario 2 to prove that due to the high performance of the algorithm, we may monitor in a mobile environment without the mobility support.

The emulator was running on top of an Amazon EC2 t2.large instance and Ubuntu 16.04.
Versions in use were Docker 1.12.1, NS3.25 and Go 1.6.2.
The containers were running as a base Ubuntu 16.04 LTS and IPv4.

\subsubsection{Scenario 1}

\imageref[0.45]{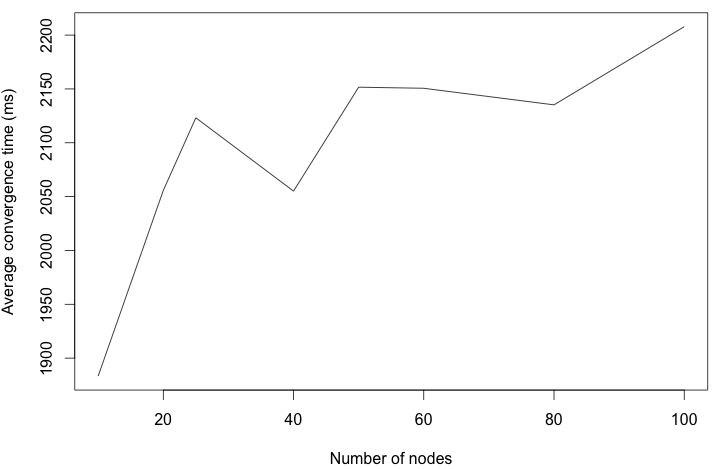}{Scenario 1 convergence results}{scenario1results}

For our first scenario, we collected the convergence time using a different root node as a starting point in each run.
We decided to use different root node selected randomly to prove that it will work independently of who the root node is.
The results are summarized in Figure \ref{fig:scenario1results}.
We can point out that there is a clear relationship between the number of nodes and the time it takes to converge.
With 50 nodes to 100 nodes, the average value seems to stabilize in around $\approx$2.15s.
About the number of packets sent, we empirically assumed that for the static environments, the number would be twice the number of nodes. This can be deducted because everyone will send their query message one time and their aggregate message one time as well.
The variability of the nodes will occur when the mobility support is added.
For $N$ nodes, the messages sent are $2*N$ for static environments.
The average message size is $\approx$151 bytes. 

\subsubsection{Scenario 2}

For the second scenario, we collected the convergence time but also the amount of observations collected by the root node at the end of each run.
The results are summarized in Table \ref{table:scenario2results}.
We can observe that the nodes converge about the same amount of time that they do in a static environment.
We observed that it would converge but without all the possible observations on the network.
And on top of that, between the more the speed of the nodes the lower the success rate would be.
By success rate, we define if the monitoring process was able to converge.
The algorithm has proved in static environments that is capable of converging really fast, even though is not using the mobility support, suggesting promising future results.

{\small
\begin{table}
  \caption{Scenario 2 results}\label{table:scenario2results}
  \vspace{-1em}
  \begin{center}
      \begin{tabular}{| l | l | l |}
      \hline
       & \textbf{Speed 2m/s} & \textbf{Speed 5m/s} \\ \hline
      \textbf{Average convergence time (ms)} & 2018.06 & 2123.63  \\ \hline
      \textbf{Average observations (\# nodes)} & 22.68 & 21.25  \\ \hline
      \textbf{Success rate} & 0.8 & 0.4  \\ \hline
      \end{tabular}
  \end{center}
  \vspace{-2em}
\end{table}
}


\section{Related Works} \label{relatedworks}

MANET monitoring has been studied 
for many objectives like their performances \cite{mehrotra2014performance}, to test them \cite{merouane2007methodology}, their security \cite{nadeem2013survey} and more recently their energetic efficiency \cite{palaniappan2015energy}.

Gossipico \cite{van2012gossip} is an algorithm to calculate the average, the sum or the count of node values in a large dynamic network.
The foundation of the algorithm is through two parts: count and beacon.
The combination of these two mechanisms 
provides the advantage for counting the nodes inside a network in an efficient and quick way.
Mobi-G \cite{stingl2014mobi} is designed for urban outdoor areas with a focus on pedestrian that moves around. 
The idea is to create the global view of an attribute incorporating all the nodes in the network.
Nevertheless the accuracy decreases for an increasing spatial network size.
On the hierarchical categorization, we can mention BlockTree \cite{stingl2013blocktree}, which is a fully decentralized location-aware monitoring mechanism for MANETs.
The idea is to divide the network in proximity-based clusters, which are arranged hierarchical.
This approach scales with the spatial network size and provides accurate results. 
In \cite{stingl2012benchmarking}, the main key points in architectural description for decentralized monitoring mechanisms are depicted. 
However, it is difficult to determine a better performer since both perform better in diverse scenarios and workloads.

\section{Conclusions} \label{conclusions}

We have presented in this paper a hybrid algorithm for monitoring decentralized networks that consists on the combination of gossip-based and hierarchical-based algorithms.
The gossip-based approach is applied to disseminate the query and the hierarchical approach is applied to aggregate the data.
Besides, with the help of a time-based hierarchical approach, the computation of a global property is achieved.
We designed a scalable and configurable testbed using NS3 and Docker, based on DOCKEMU \cite{to2015dockemu}. 
Our methodology and results seem promising for a wide set of scenarios. 

As future works, we intend to study the selection of the root node.
It could be based on location, energy, computing power and other parameters, or to be an autonomous process, proactively or reactively, or a manual process. 
Besides we plan to introduce the mobility support and enhance the testbeds to this specific cases.
We also intend to consider more complex functions in monitoring the MANETs interoperability. For this we need to define an optimal solution to propagate a more complex function through our query mechanism.

\small
\bibliographystyle{abbrv}
\bibliography{ms}


\end{document}